\documentclass{article}
\usepackage[utf8]{inputenc}
\usepackage{amsmath}
\usepackage[capitalize]{cleveref}
\usepackage{natbib}
\usepackage{color}
\newcommand{\eq}[1]{\begin{align}#1\end{align}}

\usepackage{color}
\RequirePackage[dvipsnames,usenames]{xcolor}

\title{Memory Systems, the Epistemic Arrow of Time, and the Second Law}

\author{David H. Wolpert \\
 Santa Fe Institute, 1399 Hyde Park Road, Santa Fe, NM, 87501\\
 \texttt{http://davidwolpert.weebly.com},  \texttt{david.h.wolpert@gmail.com} \\
 \emph{and} \\
{Jens Kipper}\\
{Philosophy Department, University of Rochester, NY, 14627}
 }

\date{\today}

\begin{document}

\maketitle

\begin{abstract}
The epistemic arrow of time 
is the fact that our knowledge of the past seems to be both of a different kind and more detailed than our knowledge of the future. 
Just like with the other arrows of time, it has often been speculated that the epistemic arrow 
arises due to the second law of thermodynamics.

In this paper we investigate the epistemic arrow of time, using a fully formal 
framework. We begin by defining a memory system as any
physical system whose present state can provide information about the state of the external world at some time other than the present. We then identify two types of memory systems in our universe, 
along with an important special case of the first type, which we distinguish as a third type of memory system.

We show that two of these types of memory system are time-symmetric, able to provide knowledge about both the past and the future. 
However, the third type of memory systems exploits the second law of thermodynamics in all of its instances we find in our universe.
The result is that in our universe, this type of memory system only ever provides information about the past. Finally, we argue that human memory is of this third type, completing the argument. Our analysis is indebted to prior work in Wolpert 1992, but expands and improves upon this work in several respects. 
\end{abstract}




\section{Introduction} \label{1}



It seems obvious that our knowledge of the past is of a different kind and more detailed than our knowledge of the future. It is far less obvious what explains this so-called `epistemic arrow' of time. As with the other arrows of time, the fact that the fundamental physical laws are time-symmetric presents a major obstacle to finding such an explanation.
Many philosophers and scientists have suggested explanations that appeal to the (time-asymmetric) second law of thermodynamics, or to some more fundamental facts underlying the second law \citep{Reichenbach1956, Gruenbaum1963, Horwich1987, Hawking1993, Hartle2005, schulman2005, Carroll2010, Rovelli2018, Rovelli2022, Stradis2021}. David Albert \citeyearpar{Albert2000, Albert2014, Albert2015, Albert2023} and Barry Loewer \citeyearpar{Loewer2007, Loewer2012a, Loewer2012b} have developed one such account that has been particularly influential in recent years.

Our own account is based on a formal distinction between three types of memory systems 
that occur in the physical universe, where by `memory system', we mean any kind of physical system whose present state can provide
information about the state of the external world at some time other than the present. 
On the basis of this formalism, we show that physical systems exemplifying either of the first two types can be sources of knowledge about both the past and the future. 
The epistemic arrow must therefore be grounded in the third type of memory systems. We argue that all memory systems of this type exploit a reduction of state space, which implies that the information they provide can only be of the past. 
Finally, we argue that human memory is of this third type. 
Our paper is indebted to the analysis in \citet{Wolpert1992}, but expands and improves upon it in several respects. 

The paper is structured as follows. 
In \S\ref{2}, we discuss Albert and Loewer's account. 
As we show, their explanation of the epistemic arrow relies on the doubtful idea that typically, the systems we have knowledge about had a lower entropy in the past.
We suggest that such an explanation should instead be based on the idea that the process of acquiring information involves an increase in entropy. 

In \S\ref{3}, we distinguish the three different types of memory systems we find in the physical universe, 
and present high-level examples of each type. 
In \S\ref{4} we introduce our formalism that captures the three different types of memory systems. 
We show that the third type is a special case of the first type of memory systems. 
Our investigation of how these memory systems can function reveals that one of them, namely, Type-3 memory systems, relies on the second law and is therefore time-asymmetric. 
In \S\ref{5}, we first discuss whether our account can capture the (putative) asymmetry of records. We then give reasons for thinking that human memory exemplifies Type-3 memory, which would mean that our account is suitable for explaining the epistemic arrow of time in terms of the second law of thermodynamics.
Finally, in \S\ref{6}, we spell out some remaining issues to be addressed by future research.

\section{Albert and Loewer on the asymmetry of \\records} \label{2}




Albert and Loewer's account is part of a highly ambitious project that aims to explain, among other things, all arrows of time. It begins, 
in essence, with what in the physics community has been called the argument for the ``cosmological origin of the arrow of time'' ~\citep{davies1977physics}. 
One of its key components
is what Albert and Loewer call the ``Past Hypothesis'', which is the assumption that the entropy of the very early universe was very low. 
They combine this assumption with the fact that the dynamical micro-physical laws
are deterministic and time-symmetric, and with a ``probability postulate''. 
The latter corresponds to the standard microcanonical ensemble from statistical physics, which follows from the maximum entropy principle of inference~\citep{jabr03}, and says that there is uniform probability distribution over the microstates compatible with the Past Hypothesis. 
Together, these three components determine a probability assignment to all propositions about the history of the universe. Albert
\citeyearpar{Albert2015} calls this probability assignment the `Mentaculus'.

Albert and Loewer claim that these three components also explain the ``epistemic arrow of time'',
by which they mean the fact that all records are of the past.\footnote{Many other philosophers have also appealed to an asymmetry of records---e.g., \citet{Reichenbach1956}.
}
Intuitive examples of records are impact craters, footsteps on a beach, diary entries, and memory traces in the brain. 
Albert \citeyearpar[ch.~6]{Albert2015} 
calls inference procedures that use dynamical laws to evolve macroscopic information about the present forward or backward `predictions' and `retrodictions', respectively. He states that records are those inference procedures to other times that aren't predictions or retrodictions.
A record is created when a recording device interacts with the external world---Albert calls this interaction a `measurement'. 
In typical cases, the state of the recording device will then remain stable, which allows us to draw inferences from its current state about the state of the external world at the time of the interaction.
Albert and Loewer claim that this inference requires that the recording device was in a particular state---the ``ready state''---before the interaction.\footnote{See \cite{Wolpert1992} for earlier work using the same terminology of ``predictions'' and ``retrodictions'',
making the same point about the stability of the recording device,
using the same examples,
and also highlighting the importance of a ``ready state''.}

It thus appears that to get information from a record, we need to know that the ready state obtained. 
But this seems to require another measurement, setting up a potential infinite regress. 
This regress is stopped, according to Albert and Loewer, by the Past Hypothesis, which serves as the universe's ``ultimate ready state''. By
conditioning on it, we can thus acquire knowledge of the past from records.

Of course, people had knowledge from records long before anyone had ever thought of the Past Hypothesis. 
Moreover, when we observe a record, our backward-chain of remembered measurements terminates much more recently
than 13 billion years ago, the time of the initial state of the universe. So how could the Past Hypothesis help us infer 
that our recording device was in its ready state?
As Albert explains \citeyearpar[pp.~355--358]{Albert2023}, the account isn't meant to assume that knowledge from records relies on explicit inferences from the Past Hypothesis. Rather, when we observe a record, the initial low-entropy state of the universe just makes it much more likely that the recording device was in its ready state before the time of the interaction.
He illustrates this with a half-melted block of ice sitting on the floor of a warm room. According to Albert, conditioned on the Past Hypothesis, it is likely that the block of ice was less melted several minutes in the past, and our inferences implicitly rely on this fact.
Sean Carroll \citeyearpar[p.~40]{Carroll2010} uses the example of a half-rotten egg to give a very similar account of the role of the Past Hypothesis in inferences from records. He adds that, due to the arrow of entropy, the egg's current state gives us much less information about its future states than about its past states.\footnote{
Notice that the block of ice and the rotting egg are examples of systems whose current state provides information \textit{about its own state} at a different time, rather than about the external world. 
If one considers such systems as records, then many records can give information about the future.
For example, a gas cloud with sufficient density, mass, etc. can be predicted to form a planet. 
Further examples of this type are provided by other nonlinear dynamical systems with a point attractor and an associated basin of attraction.
}

Loewer \citeyearpar{Loewer2007, Loewer2012b} generalizes this idea. He argues that, given the initial low-entropy state of the universe and the resulting arrow of entropy, information about a system's present state constrains its past states much more than it constrains its future states. The Past Hypothesis effectively imposes a tree structure on the history of the universe, with many more branches leading to the future than to the past. This implies that, typically, observations about the present state of a system give us more information about its past than about its future.

Albert and Loewer's explanation of the epistemic arrow is not fully formal. For instance, they don't provide a formal definition of records,\footnote{We will return to the question of how to define records in \S\ref{5}.
} 
or a univocal distinction between the system that contains the record and the system whose state is being recorded. 
It is nevertheless suggestive and has been highly influential, even though it has also been much criticized (cf., e.g., \citet[pp.~420--422]{Winsberg2004, Frisch2005, Frisch2007, Frisch2023, Huggett2023, Earman2006}).

Here, we would like to highlight a lacuna in their account that, to our knowledge, hasn't yet been identified. This will help us formulate a general adequacy condition for an explanation of the epistemic arrow. 

Albert and Loewer locate the source of the epistemic arrow in the entropy gradient of the objects of our knowledge, i.e., of the systems we have knowledge about. 
Their explanation thus assumes that at least typically, the entropy of physical systems we encounter is increasing.
However, the epistemic asymmetry applies to many systems whose entropy isn't increasing. 
For instance, we can have much more knowledge about what state a weather system was in 
five weeks ago than about what state it will be in five weeks from now. 
One might try to argue that such systems are not typical. But as the following considerations show, this position is untenable.

Since the appearance of the first cognitive systems on our planet, the objects of their information have almost exclusively been physical systems on Earth. 
Despite our recent efforts to learn more about the Solar System and outer space, this is still very much the case. 
The Earth system itself has remained far from thermodynamic equilibrium for a very long time. Roughly speaking, this is possible because Earth is an open system that takes in free (relatively low entropy) energy from the sun and radiates away high entropy waste heat. The entirety of the Earth system appears to be entropy-neutral---it has even been argued that its entropy has steadily decreased over the last hundreds of millions of years \citep{kleidon2009a, kleidon2009b}.
This strongly suggests that typical systems that we have information about don't exhibit an increase in entropy---there should be at least as many such systems whose entropy remains constant or is even decreasing.

At various points, Loewer adds the qualification that the relevant systems must be at least approximately thermally isolated (e.g., \citealt{Loewer2007, Loewer2020}). 
It is of course likely that most thermally isolated systems that we have knowledge about evolve towards equilibrium.
But it isn't apparent how this could be of help to their explanation of the epistemic arrow, since most of the systems that we have knowledge about aren't even approximately thermally isolated. As we just saw, the Earth system as a whole falls into this category. 

We conclude that Albert and Loewer's explanation of the epistemic arrow is at least incomplete. As we saw, a fully adequate explanation must be compatible with the fact that the entropy of many, if not most, of the systems we have knowledge about isn't increasing. 
Therefore, such an explanation shouldn't appeal to the entropy gradient of the objects of our knowledge.
The explanation we develop in what follows satisfies this condition. It is based on the idea that the epistemic arrow is due to the fact that certain processes that create information must involve a global increase in entropy.

Our investigation of the epistemic arrow of time, i.e., of the asymmetry in our knowledge of the past and of the future, doesn't assume that this arrow is constituted by an asymmetry of our knowledge of records.
In what follows, we introduce a distinction between three types of memory systems.
We then provide fully formal definitions of these three types, showing that they reflect three ways for information about the state of one system
at one time to be conveyed to the state of another system at another time.

Importantly, two of them don't yield a temporal asymmetry, and thus,
these memory systems do \textit{not} result in an epistemic arrow.
In contrast, another type of memory system we analyze involves a special initialized state (i.e., ``ready state'').
This state that allows information to be conveyed
from one moment to another is created by a process that increases global entropy.
This kind of system thus relies on the second law of thermodynamics, just like those considered by Albert and Loewer.
However, in this type of system, no assumption is made about the entropy gradient of the system it carries information about. 
Furthermore, the initialized state, too, needn't have lower entropy than the current state. Indeed, we demonstrate
that in common examples of the epistemic arrow, that initialized state has \textit{higher} entropy than the current state. 



\section{Three types of memory systems} \label{3}



A ``memory system", as we understand the term here, is any physical system whose state at the present time, $t_0$, carries information about the state of the world at time $t_1 \neq t_0$, where $t_1$ can be either in the future or in the past. By
``carry information'', we mean that
due to the
joint probability distribution of the state of the memory at $t_0$ and the state of the world at $t_1$, knowing the state of the memory at $t_0$ provides us with extra information concerning the state of the world at $t_1$, beyond our prior information about the state of the world then.



\subsection{Intuitive examples of memory systems}

To formulate this idea more carefully,
let $M$ and $W$ be the state spaces of a memory system and of the external world, respectively. Axiomatically, our probability
distributions involve the states of
$M$ and $W$ and the two times $t_0$ and
$t_1$. In addition, below we show that
in practice, the states of $M$ and / or $W$ at another time $t_2$
may play a role in memory, where
either
$t_0 < t_1 < t_2$ or $t_0 > t_1 > t_2$.

Associated with these two systems and three times, we have six jointly distributed random variables, $W_0$, $W_1$, $W_2$, $M_0$, $M_1$, and $M_2$.
Our formalizations of different types of memory
system specifies different properties of that
distribution. In this paper, we often won't discuss 
how we have come to know (sic) that the 
joint probability $P(w_0, m_0, w_1, m_1, w_2, m_2)$
over the six random variables
has those properties, or where this distribution
comes from, i.e., what physical process may
have been involved in its creation.

In what follows, we will
often be very loose with the terminology and say that we ``observe'' the state of a variable at a particular time, as shorthand for saying that we acquire some possibly noisy information about its state. Formally, such an observation involves yet more random variables, statistically coupled
with the ones described above. We ignore
such variables here.\footnote{
We don't mean to imply anything more than this by the term ``observe''. In particular, we do not imply anything involving the nature of observation in quantum mechanics.} 

For simplicity, we will speak as though 
this information we acquire
concerns the memory's present state \textit{exactly}, to infinite
precision. 
But it should be understood that in real physical systems, $M$ and $W$ will often be
elements in coarse-grainings of states in some associated phase spaces.
It is straightforward to extend
our reasoning to accommodate noisy,
imprecise information rather than such coarse-graining.


In some cases, the memory works by combining information about the
present state of the memory with information about the present state of the external world. We thus allow for
the possibility that in
addition to observing the value $m_0$, its user knows that $w_0$ falls within some particular set.
We are careful not to stipulate that the user of the memory system ``observes'' whether that is the case;
they may simply assume it. From this information about $m_0$ and possibly $w_0$,
we want to draw a probabilistic inference about the state of the external world at another time, $w_1$.

Since the memory system's present state should be relevant to the inference we draw, we require that its information
about $w_1$ varies depending on the value of $M_0$.
Intuitively, when this condition is satisfied, we can infer from the observed $m_0$ (perhaps in conjunction with $w_0$) that $M$ and $W$ interacted
sometime between $t_0$ and $t_1$, such that, in the course of this interaction, $M$ acquired information about $w_1$ and then stored it until $t_0$.


Broadly put, our taxonomy categorizes memory systems according to the kind of information they rely on. 
\emph{Type-1} memory systems involve information based on an inference from the current state of the memory system, i.e., from the value $m_0$, alone. 
\emph{Type-2} memory systems also involve information concerning the state of $m_0$, but
are only guaranteed to work when some additional conditions concerning $w_0$ are also met.
Finally, \emph{Type-3} memory systems involve information based on an inference from $m_0$ and $m_1$. 
They are a special case of Type-1 memory systems. In fact, they are the only examples of Type-1 memory systems
we know of that in the real world can accurately provide a lot of information about $w_1$, which is why we assign
them their own type. (Below we will not discuss any examples of Type-1 memory systems other than
those that are actually Type-3.)
These types of memory systems seem to capture many of the instances of memory considered
in the literature, sometime under the name of ``records''. In particular, all instances of memory we know of
that involve the second law of thermodynamics are Type-3 memory systems.

These three types of memory systems are closely related to three types of memory considered in~\cite{Wolpert1992}.
Before we formally define them, in the next subsection we present some intuitive examples of 
Type-2 and Type-3 memory systems, to compare time-symmetric memory systems (like in computers)
with time-asymmetric ones, which in practice rely on the second law and so only concern the past (like in footprints on a beach).

\subsection{How memory systems work}
\label{sec:how_memory_systems_work}

An example of a Type-2 memory system is memory in a computer. 
To keep our discussion independent of specific hardware implementations, 
we focus on abstract memory in abstract computers. 
Let $M$ be the contents of a specific  piece of Random Access Memory (RAM) that is used in a program of such a computer. 
The rest of the abstract computer---including the rest of its RAM outside of $M$, 
the program it is running, etc.---is $W$.
In such a setup, \textit{only} observing the value $m_0$ doesn't give us any information about 
$w_1$, i.e., the state of the rest of the computer at time $t_1$. The reason 
why a piece of RAM can nevertheless serve as a memory is that the entire system $M \times W$ consisting of the memory and the rest of the computer evolves deterministically in time. 
This means that we can infer something about the value of $w_1$ from an observation of $m_0$,
if we also assume (or know, via prior knowledge) a salient feature of $w_0$.
Specifically, if we know that a particular program is running on the
computer at $t_0$, then the current value of RAM, $m_0$,
can tell us the contents of some external RAM at $t_1 \ne t_0$.

It is most natural to think of such computer memory as providing information about the computer's past states. 
However, it is possible to evolve the system $M \times W$ forward in time as well as backwards, 
which means that Type-2 memory can be of the future as well as the past. 
Notice as well that our observation 
of the current state of the memory, $m_0$, can vary arbitrarily---varying that state varies what
we infer concerning $w_1$, and every value of $m_0$ provides such
an inference. On the other hand, we don't consider effects on $w_1$ of
varying the ``salient feature" concerning the state of the world external to the memory
at time $t_0$, i.e., of varying the program running on the computer. 
Instead, our inference concerning the effects of varying $m_0$
is preconditioned on $w_0$ containing a particular program, i.e., on $w_0$
falling within a particular subset of $W$.\footnote{
\citep [749--762]{Wolpert1992} discusses this kind of memory system in much more detail.
}

If $W$ is large and not fully observable, as is typical in real-life situations, then it is usually impossible to determine the precise value $w_1$ by deterministic evolution of $M \times W$.
This might suggest that Type-2 memory systems are neither very common nor very useful. However, it is compatible with our understanding of Type-2 memory systems that the inference about $w_1$ is stochastic and based on a partial observation of $w_0$---just
like with Type-1 and Type-3 memory systems. (See our formal definition below for the details.)
If one considers these kinds of cases as well, it becomes plausible that Type-2 memory systems are a very common source of
knowledge of the future. For instance, predictions of the future state of the Solar System or of the climate on Earth based on current 
observations fall into this category.

Examples of Type-3 memory are footprints on a beach, impact craters, photographic film, etc. 
Consider the case of photographic film.
Before exposure, photographic film is in a predetermined stable state, which we call its `initialized state'.
Since this state can be distinguished from any state that the film can be in after exposure, we can infer from the latter, exposed state that the film interacted in a particular way with the external world. The exposed film can thus provide us with detailed information about a past state of $W$. Since the film's state remains stable after the exposure, this state of $W$ can lie quite far in the past.

Knowledge from a (non-digital) photograph thus relies on an inference from both the present exposed state of the film, $m_0$, and its initialized state, $m_1$. This explains why photographic films are Type-3 memory systems. 
Since $m_1$ can't be directly observed at time $t_0$, the question arises how we can come to have knowledge of it. Below, we argue that this knowledge has to be based on the occurrence of a process that takes $M$ to a known state. 
Crucially, as we argue, this process of initialization must increase global entropy, which implies that $m_1$ is a past state. 
Since our argument applies to all Type-3 memory systems, this means that systems of this type can only provide information about the past.

In what follows, we develop formal definitions of the three types of memory systems just sketched, and investigate them in more detail. 
Our definitions of Type-1, Type-2, and Type-3 memory systems provide formal elaborations of Wolpert's \citeyearpar{Wolpert1992} ``b-type'', ``c-type'', and ``p-type'' systems, respectively.

\section{Formal definitions of memory systems} \label{4}



As described above, we have six jointly distributed random variables indexed by 
time, $W_0, W_1, W_2, m_0, m_1, m_2$, where 
the three associated times are {index-ordered}, i.e., either
$t_0 < t_1 < t_2$ or $t_0 > t_1 > t_2$. (We won't actually make use of $W_2$ in
what follows, except for providing some intuition.)
We are interested in forming a statistical
inference about $w_1$ based on the value $m_0$, perhaps in combination with a constraint on the
possible value of $w_0$.
We require that the inference we draw varies
depending on that value of $m_0$.
Intuitively, whenever this is the case,
we can conclude from the observed value of $m_0$ (perhaps in conjunction with an
assumed constraint on $w_0$) that $M$ and $W$ interacted
sometime between $t_0$ and $t_1$, with the interaction 
transferring  some information about
the state $w_1$ to the memory system $M$, where it resides until time $t_0$.




We can formalize the foregoing with what we call \textbf{memory systems}.
We consider three types of memory systems, which differ from one another
depending on whether the memory is based on the value $m_0$, on the value $w_0$, or on the value $m_0$ combined with some knowledge about how
the laws of physics arise in the joint dynamics of
$M \times W$. 

In the rest of this paper, for simplicity we consider
the case where all state spaces
are countable, e.g., due to coarse-graining.
The extension to uncountable spaces is straightforward. In addition, we
write the indicator function
as $\delta(.)$. So for any event $A$ in the implicit underlying
probability measure space, $\delta(A)$ equals
$1 / 0$ depending on whether $A$ is true/false.

In the next subsection, \cref{sec:restricted},
we begin by introducing 
a variant of some standard information-theoretic
definitions. These will play a central role
in our fully formal definitions of those three types
of memory systems, which we present in 
\cref{sec:formal_defs}.

\subsection{Restricted mutual information}
\label{sec:restricted}

In general, whether the state $m_0$ provides a memory 
about the state $w_1$
will depend on certain conditions concerning the joint value 
$(m_0, w_0)$ being met. 
%
Accordingly, 
our definitions will involve
statements of the form, ``If condition $\mathcal{C}$ concerning $(m_0, w_0)$ is 
met, then the following mutual information will be high". 
We will not model how the user of the memory system
does (or doesn't) come to know whether condition $\mathcal{C}$
is met. Often it will be background knowledge, over and beyond the 
background knowledge that determines the joint distribution
$P(m_0, w_0, m_1, w_1, m_2, w_2)$. 

To illustrate this, consider again
the example of a computer memory described above. In that example,
$M$ is (the state of) 
part of a computer's RAM, and $W$ is (the state of) 
the rest of the computer, including in particular the
rest of the RAM. $P(.)$ depends on the dynamics of 
the entire computer, as usual. In this example, condition $\mathcal{C}$ is the
knowledge that some specific program
is currently executing in $W$, the rest of the computer
outside of the part of the RAM constituting $M$. 
It is the precise form of that program which,
combined with the current state of the part of the RAM constituting $M$,
provides information concerning the state of the rest of the RAM at
some other time. Note that in this example the constraint doesn't
specify $w_0$ \textit{in toto}; many degrees of freedom of the computer
are free to vary.

Intuitively, knowledge that $\mathcal{C}$ holds
is a second, different kind of ``observation", in addition to
the observation of the precise current state of the computer
memory in question. The difference between the two types
of observations is that we will be considering
the effect on what we can infer about $w_1$ by varying over
the states $m_0$, while we will not consider varying over 
whether $\mathcal{C}$ holds. Again returning to the example of a computer,
we distinguish the observation of the part of the RAM that
comprises $M$ from the ``observation" of what program is running on
the rest of the computer. We are interested in how varying the former
leads to different conclusions concerning the state of the external RAM
at some other time. In contrast, we are not concerned with the effects of
varying the program.
 
To formalize this distinction, for any jointly distributed
pair of random variables $A, B$ taking values $a, b$ respectively,
let $\mathcal{C}$ be some set of joint values $(a, b)$.
Define $C$ to be the indicator function specifying whether 
$(a, b) \in \mathcal{C}$. So $C$ is a
$0/1$-valued random variable, jointly distributed with our other random variables. Indicate the joint distribution as $P(a, b, c)$, where
$c$ is the value of $C$.
Then we can define the random variable, 
\eq{
I_c(A ; B) 
:= -\sum_{a, b} P(a, b | c)  \left[ \ln P(a | c)
- \ln P(a | b, c)\right]
\label{eq:fixed_mutual_information}
}
Intuitively, $I_c(A ; B) $ is the value of the mutual information
between $A$ and $B$, evaluated only over those $(a, b)$ pairs
where condition $\mathcal{C}$ does / doesn't hold, as specified by
the value of $c$. Note that 
$I_c(A ; B)$ isn't the 
same as the mutual information between $A$ and $B$ conditioned
on $c$,
\eq{
I(A ; B | C) = -\sum_{a, b, c} P(c) P(a, b | c)  \left[ \ln P(a | c)
- \ln P(a | b, c)\right]
}
Indeed,
$I(A ; B | C)$ is the expectation under $P(c)$ of
$I_c(A ; B) $. 

We can illustrate this definition by returning to the example
where $M$ is a part of a computer, while the program running
in the computer is stored in some other part of the RAM which
is (part of) $W$. In this example, $c = 1$ iff
the joint state of a RAM in a computer and the program stored
in the rest of the computer fulfills some condition.

We will refer to $I_c(A ; B)$ for $c =1$ as the ($\mathcal{C}$-)\textbf{restricted} mutual information
between $A$ and $B$, 
We will write it as $I_{\mathcal{C}}(A ; B)$, with
the value $c = 1$ implicit.

Memory systems are defined in terms of \textit{sufficient}
conditions for information concerning the external
world at one time to be conveyed to the memory system
at another time, and we make no claims about \textit{necessary and sufficient} conditions.
For this reason, in this paper we are interested
in restricted mutual information rather than conditional mutual information, with 
$C = 1$ for different choices of $\mathcal{C}$ being the sufficient conditions.

As an aside, note that we can define variants of entropy
and conditional entropy that are analogous to $I_c(A ; B)$:
\eq{
H_{c}(A) &:= -\sum_a P(a | c) \ln P(a | c) \\ 
H_c(A | B) &:=  -\sum_{a.b} P(a, b | c) \ln P(a | b, c)
}
where as before, $c \in C$ is a 0-1 valued random variable specifying
whether condition $\mathcal{C}$ holds. 
For any such random variable $C$
and either value $c$ of that random variable,
\eq{
I_c(A ; B) = H_{c}(A) - H_c(A | B)
}

Paralleling our convention for restricted mutual 
information, we will sometimes write
the two types of restricted entropy evaluated for $c=1$ as $H_{\mathcal{C}}(A)$
and $H_{\mathcal{C}}(A | B)$, respectively. So in particular,
\eq{
I_{\mathcal{C}}(A ; B) = H_{\mathcal{C}}(A) - H_{\mathcal{C}}(A | B)
}
in direct analogy to the relation among (non-restricted) entropy, conditional entropy, and
mutual information.

As a point of notation, we will often write something like ``$a \in \mathcal{C} \,$'' inside a probability distribution 
as shorthand for the event that the value of the associated random
variable $C = 1$. Similarly, we will write $I_{a \in \mathcal{C}}( \ldots)$ as shorthand for a $\mathcal{C}$-restricted
mutual information where the variable $a$ lies in the set $\mathcal{C}$. Furthermore, 
 let $d \in D$ be some random variable. Rather than write 
``for all $(a, b) \in \mathcal{C}, P(d \,|\, a, b)$ obeys ..." it will be convenient to write
``$P_{(a, b) \in \mathcal{C}}(d \,|\, a, b)$ obeys ...''.

\subsection{The three types of memory systems}
\label{sec:formal_defs}

$ $

\noindent \textbf{Definition 1:}    A \textbf{Type-1} memory is any stochastic process over the space $M \times W$ 
where there is some set $M^*$ such that
$I_{m_0 \in M^*}(W_1 ; M_0)$, is large.
        
        $ $
    
\noindent \textbf{Definition 2:}  A \textbf{Type-2} memory
    is any stochastic process over the space $M \times W$ where there is some set $W^*$ such that
    $I_{w_0 \in W^*}(W_1 ; M_0)$ is large.

        $ $


    
    
\noindent \textbf{Definition 3:}  
A \textbf{Type-3} memory is any stochastic process over the space $M \times W$ where:
    \begin{enumerate}
        \item 
        \label{item:3.1} There is an $m^\dagger \in M$ and a set $M^*$ such that
        $I_{m_1 = m^\dagger, m_0 \in M^*}(W_1 ; M_0)$ is large.

        
\item There is a set $M' \subseteq M$ such that for all $m_0\in M^*$:
        \begin{enumerate}
          \item $P(m_2 \in M' \,\vert\, m_0)$ is
            close to $1$. 
\label{item:3.2.i}
        
        \item $P(m_1 \,\vert\, m_2, m_0)$ is a highly peaked 
        distribution about $m_1 = m^\dagger$, for all $m_2 \in M'$.
\label{item:3.2.ii}
        
        \item $w_1$ is conditionally independent of $m_2$,
        given $m_0$ and given that $m_1 = m^\dagger$. In other words,
\eq{
P(w_1 \,\vert\, m_0, m_1 = m^\dagger, m_2)  = P(w_1 \,\vert\, m_0, m_1 = m^\dagger) \nonumber
}
\label{item:3.2.iii}
    \end{enumerate}
\label{item:3.2}
\end{enumerate}

\cref{item:3.1} of the
definition of Type-3 memory systems says that if we believe for some reason
that the memory is in the initialized state $m^\dagger$ 
at $t_1$, and if $m_0 \in M^*$, then knowing the
precise value $m_0$ provides a lot of information about $w_1$. Intuitively, knowing both
that the system was in  $m^\dagger$ at $t_1$ and that $m_0 \in M^*$, we can conclude that
$W$ must have interacted with $M$ between $t_1$ and $t_0$, with the precise relationship between
$m^\dagger$ and $m_0$ providing information about the state of $W$ before that interaction started,
at $t_1$. \cref{item:3.1} says that we have reason to belief that $m_1$ does in fact equal $m^\dagger$,
and so we can use $m_0$ to make an inference about $m_1$ this way.

As established in Lemma\,1 below, \cref{item:3.2.i} and \cref{item:3.2.ii} of Def.\,3 then 
provides a set of properties of the joint probability distribution
that justify that belief concerning $m_1$, the state of the memory at $t_1$, given
only the fact that the present state of the memory system is in $M^*$. (\cref{item:3.2.iii} is a simplifying assumption,
made for expository convenience.)

Thm.\,1 below then uses Lemma 1 to show that when the conditions for a Type-3 memory system hold,
$I_{m_1 = m^\dagger, m_0 \in M^*}(W_1 ; M_0)$ is large. This proves that Type-3 memory systems
are a special case of Type-1 memory systems. In fact, as also discussed below, Type-3 memory
systems are an especially important special case of Type-1 memory system, since they can be considered as a formalization of the primary type of memory system that is 
considered to be a ``record of the past'' in previous literature on the epistemic arrow of time. 
The second law of thermodynamics
seems to play a crucial role in allowing the properties the conditions defining Type-3 memory systems
(in particular, \cref{item:3.2.ii}) to occur in the real world. In contrast, the second law doesn't arise at all in Type-2 memory systems. 

$ $

\noindent \textbf{Lemma 1:}
For a Type-3 memory,
\begin{enumerate}
\item 
\label{item:lemma_1.i}
For any $m_0 \in M^*$ and any $w_1$,
\eq{
P(w_1 \,\vert\, m_0) \simeq
    P(w_1 \,\vert\, m_0, m_1 = m^\dagger)
}
and since this holds for all $m_0 \in M^*$,
\eq{
P(w_1 \,\vert\, m_0, m_0 \in M^*) &\simeq P(w_1 \,\vert\, m_0, m_1 = m^\dagger, m_0 \in M^*)
\label{eq:9}
}

\item 
\label{item:lemma_1.ii}
For any $m_1$,
\eq{
P(m_1 | m_0 \in M^*) &\simeq \delta(m_1, m^\dagger)
}

\item 
\label{item:lemma_1.iii}
For any $m_0$,
\eq{
P(m_0 | m_0 \in M^*) 
     &\simeq P(m_0 | m_1 = m^\dagger, m_0 \in M^*)
\label{eq:11}
}

\item 
\label{item:lemma_1.iv}
For any $w_1$,
\eq{
P(w_1 \,\vert\, m_0 \in M^*) &\simeq P(w_1 \,\vert\, m_1 = m^\dagger, m_0 \in M^*)
}

\end{enumerate}

\noindent \textbf{Proof:}
For any $m_0 \in M^*$ in a Type-3 memory, we can expand
        \eq{
          P(w_1 \,\vert\, m_0) &= \sum_{m_1, m_2} P(m_2 \,\vert\, m_0)
                        P(m_1 \,\vert\, m_2, m_0) P(w_1 \,\vert\, m_0, m_1, m_2) \\
             &\simeq \sum_{m_1, m_2} 
             \frac{P(m_2 \,\vert\, m_0) 
               \delta(m_2 \in M')}
             {\sum_{m_2}P(m_2 \,\vert\, m_0) 
               \delta(m_2 \in M')}
               P(m_1 \,\vert\, m_2, m_0) P(w_1 \,\vert\, m_0, m_1, m_2) \\
            &= \sum_{m_1, m_2} 
            \frac{P(m_2 \,\vert\, m_0) }
             {\sum_{m_2}P(m_2 \,\vert\, m_0) 
               \delta(m_2 \in M')}
               \delta(m_2 \in M')P(m_1 \,\vert\, m_2, m_0) P(w_1 \,\vert\, m_0, m_1, m_2) \\
            &\simeq \sum_{m_1, m_2} 
            \frac{P(m_2 \,\vert\, m_0) }
             {\sum_{m_2}P(m_2 \,\vert\, m_0) 
               \delta(m_2 \in M')}
               \delta(m_2 \in M')\delta(m_1 - m^\dagger) P(w_1 \,\vert\, m_0, m_1, m_2) \\
                    &=\sum_{m_2} P(m_2 \,\vert\, m_0) 
                     P(w_1 \,\vert\, m_0, m_1 = m^\dagger, m_2) \\
                    &=\sum_{m_2} P(m_2 \,\vert\, m_0) 
                          P(w_1 \,\vert\, m_0, m_1 = m^\dagger) \\
                    &= P(w_1 \,\vert\, m_0, m_1 = m^\dagger)
        \label{eq:4}
            }
        where the second line uses \cref{item:3.2.i},
        of the definition of Type-3 memory systems, the fourth line uses \cref{item:3.2.ii} and then
        the sixth line uses \cref{item:3.2.iii}. This establishes Lemma 1(1).

Next, expand 
\eq{
P(m_1 | m_0 \in M^*) &= \sum_{m_2} P(m_1 | m_0 \in M^*, m_2) P(m_2 | m_0 \in M^*)  \\
    &\simeq \sum_{m_2} P(m_1 | m_0 \in M^*, m_2) P(m_2 | m_0 \in M^*)\delta(m_2 \in M^\dagger)  \\
 &\simeq \sum_{m_2} \delta(m_1, m^\dagger) P(m_2 | m_0 \in M^*)\delta(m_2 \in M^\dagger)  \\
  &=  \delta(m_1, m^\dagger)
}
where the second line uses \cref{item:3.2.i} 
of the definition of Type-3 memory systems,
and the third line uses \cref{item:3.2.iii}.
This establishes Lemma 1(2).

Next, use Lemma 1(2) to expand
\eq{
P(m_0 | m_0 \in M^*) &= \sum_{m_1} P(m_0 | m_0 \in M^*, m_1) P(m_1 | m_0 \in M^*)
        \\
     &\simeq \sum_{m_1} P(m_0 | m_0 \in M^*, m_1) \delta(m_1, m^\dagger) \\
     &= P(m_0 | m_0 \in M^*, m_1 = m^\dagger)
}
This establishes Lemma 1(3).

Finally, apply $\sum_{m_0} P(m_0 | m_0 \in M^*)$ to both
sides of \cref{eq:9}, and then use \cref{eq:11}
to replace $P(m_0 | m_0 \in M^*)$ in the right-hand
sum. This establishes Lemma 1(4).

\noindent \textbf{QED}

$ $

We can use Lemma 1 to derive the following result, and thereby prove that systems obeying
the four properties of Type-3 memory systems are in fact a special case of Type-1 memory systems,
as claimed above:

$ $

\noindent \textbf{Theorem 1:}
$I_{m_0 \in M^*}(W_1 ; M_0)$ is large in any Type-3 memory
system.

$ $

\noindent \textbf{Proof:} Using Lemma 1(1) twice allows us to expand

\eq{
I_{m_0 \in M^*}(W_1 ; m_0) 
    &= \;-\!\!\sum_{m_0, w_1} P(m_0, w_1 | m_0 \in M^*)  
    \bigg[ \ln P(w_1 | m_0 \in M^*) - \ln P(w_1 | m_0, m_0 \in M^*)\bigg] 
 }
 \eq{
&\simeq \;-\!\!\sum_{m_0, w_1} P(m_0, w_1 | m_0 \in M^*)  \bigg[ \ln P(w_1 | m_0 \in M^*) - \ln P(w_1 | m_0, m_1 =     
    m^\dagger,m_0 \in M^*)\bigg] \\
&\simeq \;-\!\!\sum_{m_0, w_1} P(m_0 | m_0 \in M^*) 
    P(w_1 | m_0, m_1 = m^\dagger,m_0 \in M^*) \nonumber \\
&\qquad\qquad \qquad \times \bigg[ \ln P(w_1 | m_0 \in M^*) - \ln P(w_1 | m_0, m_1 = m^\dagger,m_0 \in M^*)\bigg]            
\label{eq:18} 
}

\noindent Next, we can use Lemma 1(3) and then Lemma 1(4) to 
approximate \cref{eq:18} as 
\eq{
I_{m_0 \in M^*}(W_1 ; m_0) 
    &\simeq \sum_{m_0, w_1} P(m_0 | m_1 = m^\dagger, m_0 \in M^*) 
    P(w_1 | m_0, m_1 = m^\dagger,m_0 \in M^*)  \nonumber \\
&\qquad\qquad \qquad \times \bigg[ \ln P(w_1 | m_0 \in M^*) - \ln P(w_1 | m_0, m_1 = m^\dagger,m_0 \in M^*)\bigg]    \\
  &\simeq \sum_{m_0, w_1} P(m_0 | m_0 \in M^*, m_1 = m^\dagger) 
    P(w_1 | m_0, m_1 = m^\dagger,m_0 \in M^*)  \nonumber \\
&\qquad\qquad \qquad \times \bigg[ \ln P(w_1 | m_1 = m^\dagger,m_0 \in M^*) - \ln P(w_1 | m_0, m_1 = m^\dagger,m_0 \in M^*)\bigg] \\
  &= \sum_{m_0, w_1} P(m_0, w_1 | m_1 = m^\dagger, m_0 \in M^*)     \nonumber \\
&\qquad\qquad \qquad \times \bigg[ \ln P(w_1 | m_1 = m^\dagger,m_0 \in M^*) - \ln P(w_1 | m_0, m_1 = m^\dagger,m_0 \in M^*)\bigg] \\
  &= I_{m_1 = m^\dagger,m_0 \in M^*}(W_1; M_0)
}

\noindent Finally, plugging in \cref{item:3.1} of
the definition of memory systems, we conclude that
$I_{m_0 \in M^*}(W_1 ; m_0)$ is large. 

\noindent \textbf{QED}







$ $

Thm.\,1 establishes that in a Type-3 memory system, so long as $m_0 \in M^*$, the precise state $m_0$
is informative about the state $w_1$. So whenever that condition is met, the current state of the memory
system $M$ is a \textit{memory}
of $w_1$, the state of the external world at $t_1$,
in the sense described in preceding sections.

\subsection{Illustrations of our formal definitions}
\label{sec:formal_defs_examples}

In this subsection we illustrate real-world examples of Type-2 and Type-3 memory systems,
to compare the formal definitions of time-symmetric and time-asymmetric memory systems. We can illustrate the 
definition of Type-2 memory systems using the above example of a computer memory. Recall that in that example,
$M$ is one part of the RAM of the computer, while
$W$ is the rest of the RAM, including in particular the part of the RAM that contains  
the program currently running on the computer. More precisely, write the state space
of the computer as $Z = (M, Z_2, Z_3)$, where $z_2 \in Z_2$ specifies the particular program currently
running (i.e., a particular interval of the coding segment of the computer), \textit{except} for one
variable in that program, namely $m \in M$. $Z_3$ is then
the rest of the RAM and other variables in the computer whose value isn't involved in specifying
the program. 

So in this computer memory example, $W$ is $(Z_2, Z_3)$. However, it is 
convenient to parameterize elements of
$W$ by their value of $Z_2$, coarse-graining over possible values of $Z_3$.
In particular, $W^*$ is all states of $(Z_2, Z_3)$
where $z_2$ contains a particular program $z_2$, a program
that allows inference from the current state of
the memory, $m_0$, about the past and / or future of the variable $m$. Note that such
inference relies on the fact that the typical values of $z_3$ have no effect on the dynamics
of $(m, z_2)$.

More generally, in many Type-2 memory systems, $M \times W$
is a closed system, isolated from the rest of the physical universe during the interval $[t_0, t_1]$.
In such a case, since the laws of physics are deterministic and invertible in any closed system,
the joint dynamics of $M \times W$ 
is deterministic during $[t_0, t_1]$. Type-2 memory systems with this
feature can result in almost {perfect} memory, as described in \cref{sec:how_memory_systems_work}.


We can illustrate the different parts of the definition
of Type-3 memory systems with the example of a line of footprints across a beach.
In this example,
$M$ is the set of all versions of the \textit{pattern} on the surface of the beach---smoothed, with a single line of footprints,
churned by many people walking across it, etc. 
$M^*$ is the set of all versions of the (pattern on the surface of a) beach that are completely smooth, having been
swept by ocean waves, during a high tide, with the possible exception that it
has a very clearly defined line of footprints. $M'$ is all 
versions of the beach that are not in some unusual state that would prevent the beach
from being swept smooth. $m^\dagger$, the ``initialized state'',
is the beach right after it has been smoothed by ocean waves.\footnote{N.b., strictly speaking, $m^\dagger$ isn't a single state, but a set of very similar states. To simplify the exposition, we will often treat a set of very similar states as though
they were a single state, as was also done in the example above of a computer memory.} In contrast,
$W$ is the set of all other systems on the surface of the Earth that could conceivably interact with the surface of
the beach some time in the interval between $t_0$ and $t_2$.

\cref{item:3.1} reflects the fact that if we know both
that the beach surface was smooth at $t_1$ and that it currently is smooth except for a single line
of footprints, then we can conclude that a person must have walked across the beach 
some time between $t_1$ and $t_0$, with the precise pattern of those footprints
providing information about that walk. 

 \cref{item:3.2.i} of the definition of Type-3 memory systems then tells us so long as the current pattern on the beach is 
a single line of footprints, we have no reason to suppose that the surface of the beach was in some
unusual state that could not be wiped smooth just before the most recent high tide.

\cref{item:3.2.ii} 
of the definition of Type-3 memory is enforced by the second law of thermodynamics. More precisely, the collapsing of the state space of $M$
described in  \cref{item:3.2.ii} involves coupling $M$ with some third system, $K$.
The second law drives an irreversible process that increases total entropy in $M \times K$, while at the same time
collapsing $M$ from the subset $m_2 \in M'$ down to the precise value $m_1 = m^\dagger$.\footnote{This is related to
what was called ``external initialization'' in~\cite{Wolpert1992}.}

Concretely, a beach has just been initialized as $m^\dagger$ when it has just been smoothed by the ocean waves driven by the tide.
$K$ is those ocean waves, lapping the beach during this re-initialization of the state of the beach.
Projected down to the states of the beach,
that smoothing of the beach by ocean waves is a non-invertible process, driven by the second law.
This reliance on the second law of course is 
precisely why this example of a Type-3 memory system
is time-asymmetric.\footnote{As noted above, \cref{item:3.2.iii} is assumed simply for expository convenience,
and clearly holds for this example of a beach.}

Note that just like with Type-2 memory systems, with Type-3 memory systems there is an
implicit assumption that $W$ is a minuscule portion of the full physical
universe.\footnote{More precisely, we assume that the probability that variables in the physical universe
that lie outside of $W$ are in a state that would cause them to interfere with our inference
is effectively $0$.} Furthermore, it is implicitly
assumed that the dynamics of those degrees of freedom of $W$ we are concerned with
are effectively isolated from that rest of the universe (aside from the possible interaction
with a system $K$). 
This assumption implies, for instance, that the sand on the beach was not manipulated by powerful aliens to make it appear as though people had walked over a smooth beach.

Note also that the fact that the distribution over $m$ at $t_1$, the end of
the initialization process, is (almost) a delta function about $m^\dagger$
means that the distribution over $M$ at that time, when it's in
its initialized state, has \textit{low entropy}. It is
the distribution over the joint state, $M \times K$, 
whose entropy increases in the initialization of $M$.

A flash drive is another example of Type-3 memory that provides an even more graphic illustration of how the initialized, ready state
of $M$ can have low entropy.
Here, 
$M = (Z_1, Z_2)$, where $Z_1$ is the contents of the flash drive's binary memory, and $Z_2$ is other
attributes of the physical flash drive, in particular whether it has physical
damage (e.g., puncture holes in the flash drive's casing). 
$M^* = M' $ is all joint states in $ (Z_1, Z_2)$ where ($Z_2$ has a value indicating that) the flash drive is undamaged.
$m^\dagger$ is the ``wiped clean'', all-$0$'s joint state of the flash drive's entire memory, i.e., of $Z_1$. 

The important thing to note is that this wiped-clean state where the bits are all $0$'s with probability $1$ is \textit{minimal}
entropy. It is produced by coupling the flash drive with an external, electronic initializing system, $K$,
in a ``wiping clean'' process of the contents of the flash drive.
That initialization process relies on the second law of thermodynamics to increase the joint entropy
of the flash drive \textit{and the electronic initializing system}.
So just like the beach was wiped smooth by the action of waves during a high tide, which increased
the joint entropy of the waves and the beach while reducing the marginal entropy of just the beach,
the flash drive was wiped clean by action of the electronic initializing system, which increased
the joint entropy of the initializing system and the flash drive's bits while reducing the marginal entropy of just the 
flash drive's bits.

As an alternative, we could reformulate these examples of Type-3 memory systems
not to involve an external system $K$. We would do this by ``folding $K$ in''
to the definition of $M$. 
In the example of a beach surface memory system, this 
would mean redefining $M$ to be the joint state of
pattern on the surface of the beach \textit{and the precise physical state of the ocean lapping that beach}. 

Finally, we note that it is straightforward to formalize 
other examples of memory systems  considered in the literature (in particular,~\citep{Wolpert1992})
as Type-3 memory systems. 
To illustrate this, consider the example of an image on a chemical photographic film in an instant camera. $M$ is the possible
patterns on the surface of the film; $M^*$ is all such patterns aside from those that indicate
the camera holding the film was incorrectly exposed to the outside world, e.g., resulting in a fogged
image on the surface of the film. $m^\dagger$ is the initialized state of the film, with no image, 
before exposure of any sort. It has low entropy, and is formed in an entropy-increasing chemical initialization process that
involves some external set of chemicals, $K$.  $W$ is an external photon field,
which will result in an image being made some time between $t_1$ and $t_0$
if the camera exposes the film correctly, i.e., if $m_0 \in M^*$.

%


\subsection{Discussion of our formal definitions}
\label{sec:formal_defs_discussion}


In this subsection we briefly discuss some aspects of the formal definitions of the various types of memory system. 


First, note that while there is no need to do so here, we could replace
phrases like ``$I_{m_0 \in M^*}(W_1 ; M_0)$ is large'' 
with more formal expressions. For example, suppose that
both $|M^*|$ and $|W|$, the number of states in $M^*$ and in $W$, respectively, are finite. 
Then we could replace that phrase
by saying that $I_{m_0 \in M^*}(W_1 ; M_0)$
is close to $\min (\ln |M^*|, \ln |W|)$, its maximum possible
value.

Note also that in Type-1 and Type-3 memory
systems, we allow the possibility that we can know the value $m_0$ even if it is
 outside of $M^*$.
We even allow for the possibility that
there would be nonzero mutual information between the value of $m_0$ and that of $w_1$ for $m_0 \not\in M^*$. However, our analysis concerns what happens when $m_0 \in M^*$. (\textit{Mutatis mutandi}
for values of $w_0$ being outside of $W^*$ in the case of Type-2
memory systems.)
 

In real-world Type-3 memory systems, often $m$ will not change in $[t_2, t_0]$ except at the time of its interaction with $W$. 
While we do not require this, it has 
the practical advantage that 
it simplifies the calculation by the memory's user
of the relationship between the value of $w_1$ and $m_0$. It also means that we
don't need to be precise about when the times $t_1$ and $t_2$ are.

It is important to realize that the system $K$ in Type-3 memory systems, which couples with $M$
in an entropy-increasing process to send $M'$ to $m^\dagger$, doesn't explicitly occur in
the definition of Type-3 memory systems. Rather it arises \textit{in practice}, as part of the underlying
process that enforces the requirement in \cref{item:3.2.i} that the conditional
distribution $P(m_1 \,\vert\, m_2, m_0)$ is peaked about $m_1 = m^\dagger$. 
In turn, that requirement is only relevant under the supposition
that $m_0 \in M^*$ and $m_2 \in M'$.

There are many important ways that the analysis in this paper extends beyond / modifies
the analysis in \cite{Wolpert1992}, which was written 
before the revolutionary
advances of the last two decades of stochastic thermodynamics. Like all considerations of the
thermodynamics of computation at the time, it was based on semi-formal reasoning, grounded in
equilibrium statistical physics. However, computers are
actually very far from thermal equilibrium, with the result that the understanding of
the relationship between logical and thermodynamic irreversibility at the end of the twentieth century 
and its implications for the thermodynamics of computation was mistaken.
Our paper doesn't rely on that mistaken earlier understanding, and is fully consistent with our
modern understanding of statistical physics.
(Cf.~\citep{sagawa2014thermodynamic,wolpert2019stochastic}
and references therein for an introduction to the modern understanding  of the relationship between logical and
thermodynamic irreversibility.)

Another important feature of \cite{Wolpert1992} is its repeated invocation of the Maxent principle of
Jaynesian inference. In this paper we do not use Maxent. Indeed, we are careful to make no 
arguments about how it is that the user of a memory system may arrive at the probability
distributions they are using. 
In particular, it is worth noting that in this paper, we make no \textit{a priori} assumption 
that $P(m_0, m_1, w_0, w_1)$
has full support \citep [cf.] [fn. 9] {Wolpert1992}.


\section{Memory systems, records, and the epistemic arrow} \label{5}
\label{sec:epistemic_arrow}



Of the three types of memory system we have considered, 
Type-3 systems are the only ones that, at least in all of their instances we know of in our physical universe, are time-asymmetric, in that they can only provide information about the past.
As we explained, Type-3 memory systems rely on the second law, in that they exploit the fact that an increase in global entropy reliably takes the (local) memory system to its initialized state, which is a known state at $t_1$.

While we have not proven it,
we note that in practice, the only way the need for the second law can be circumvented without major sacrifice in
the accuracy of the memory is if 
%
we have detailed knowledge of 
those ``dynamically relevant'' degrees of freedom in the present state of $W$ that (perhaps together with the precise state of $M$)
determine the dynamics of $M$. In practice, as in the computer example of Type-2 memory systems,
we in fact have a way to (almost) deterministically calculate the 
joint dynamics of $M \times W$. 

Note that these requirements do not preclude the possibility
that $W$ is extraordinarily large.\footnote{For example, a
modern cloud computer system has tens of thousands of servers, each 
having $\sim 10^{15}$ (?) dynamically relevant degrees of
freedom. So setting $M$ to be part of the memory of just one of those servers,
$|W|$ is on the order of Avogadro's number. Yet such computer systems are
examples of Type-2 memory systems.}
%
However, to run a Type-2 memory 
system with a large $W$ seems to require 
a huge number of energy barriers keeping
trajectories of $M \times Z_2$ well-separated
during the evolution of the joint system, with high probability, i.e., such systems use a huge amount
of error correction. (This is certainly true in cloud computers.) Systems with this property seem to only
arise with careful engineering by humans. In contrast, memory systems like footprints on a beach do
not rely on anything close to that number of energy barriers, allowing the stochastic process governing
the dynamics of microstate trajectories to spread out more readily. This may be why they
can occur in systems that are not artificially constructed. (See discussion of 
the Past Hypothesis in \cref{6}.)

In what follows, we discuss whether Type-3 memory systems might correspond to records. After this, we
argue that human memory is plausibly Type-3, which would mean that our analysis is suitable for explaining the epistemic arrow of time.

Common examples of records, such as impact craters, footsteps on the beach, and photographic film, are Type-3. Furthermore, Albert and Loewer claim that records require a ready state, and the initialized state formalized
in our definition of Type-3 memory systems as $m^\dagger$ is such a ready state.
Does this mean that Type-3 memory systems can be interpreted as a formalization of records? 
In the absence of a precise definition of records, this question is difficult to answer. 
We believe that for this interpretation to work, one needs to assume that it is true by definition that records rely on an initialized state---otherwise, we don't see a clear way to distinguish records from Type-2 memory systems.
If this assumption is made,
then our analysis (which in turn builds on the work in~\cite{Wolpert1992}, as
described above) might provide a new basis for understanding Albert and Loewer's claim
that the epistemic arrow is constituted by the temporal asymmetry of records which avoids the problematic aspects of their argument (cf.
\cref{2}). 

At present, the physical details of how the human brain stores information are largely unknown. 
This makes it difficult to determine what type of memory system the human brain represents.
Nevertheless, there are reasons to think that human memory is Type-3.
First, there is the simple fact that human memory only provides information about the past. 
Since Type-3 memory systems are the only memory systems that exhibit this kind of temporal asymmetry, this suggests that human memory are Type-3.
Second, human memory in the primary sense resides in the brain---we might call this `internal memory'. But humans also remember things indirectly by means of external devices, such as photographs, books, or digital storage media---we might call this `external memory'. 
External memory, at least if it concerns information about events occurring outside of computers, is typically Type-3. 
(Our discussion in \ref{sec:formal_defs_examples} demonstrated this for some such systems, namely photographs and flash drives.)
This makes it possible for such memory to store very detailed information. 
Internal memory, too, often provides us with highly detailed information about specific events. 
An important aspect of the psychological arrow of time is that we experience the future as ``open'' and the past as ``fixed''.\footnote{Cf. \citep [pp. 776--778]{Wolpert1992} for further discussion of the relation between this aspect of the psychological arrow and the epistemic arrow. 
}
It is plausible that the fact that we have such detailed memories of the past is at least part of the cause of this apparent openness of the future and fixity of the past.
The fact that internal memory can provide such detailed information supports the idea that it is Type-3.
If this is the case, then our analysis is suitable for explaining how the epistemic arrow arises from the second law of thermodynamics.

\section{Future work and open issues}
\label{6}

There are many avenues for investigation that the analysis in this paper highlights but does not address.

In this paper we consider three types of memory systems, which are the three types of memory system we can find examples of in the real, physical world.
We provide no proof that no other type of memory system is possible. One obvious
avenue for future work is to investigate this issue further.

We show how, due to the second law,
there can be Type-3 memory systems of the past.
We also argue (semi-formally) that the human brain involves
such types of memory.
Based on our discussion, we consider it plausible that Type-3 memories cannot be of the future. 
In essence, this is because we don't see a potential mechanism that could play the role the second law of thermodynamics plays in such putative Type-3 memories of the future.
But we provide no formal proof that Type-3 memory systems can only be of the past. This issue will thus have to be left for future research.


Another important issue builds from the discussion at the end of \cref{sec:formal_defs_discussion}: how
exactly is it that the user of the memory comes to ``know'' the joint
distribution in the first place? Does acquiring that knowledge itself rely on memory, of past observations
of the physical world? This is an extremely subtle issue, which ultimately requires engaging with the
formal impossibility of inductive inference~\citep{adam2019no,enwiki:1170710847,wolpert2023implications}.
\textit{If} the joint probability distributions of $M \times W$ at multiple moments in time
has the structure of a Type-3 memory system formally defined in \cref{sec:formal_defs}, then the 
relevant mutual information can in principle be exploited. Moreover, sidestepping the problem of inductive
inference~\citep{wolpert2023implications}, speaking purely as empirical scientists, it seems likely that natural selection 
has guided (the genes encoding) our biological memories to assume those distributions, in order to
increase our biological fitness. But in this paper, we do not grapple with these issues.

Yet another deep problem involves the asymmetry of the second law, which appears to be fundamental
to (the asymmetry of Type-3 memory and therefore) the asymmetry of human memory.
We are sympathetic to the idea of grounding the second law in the ``Past Hypothesis'. Stated
carefully, this hypothesis has three parts. First, it (implicitly) models the dynamics of the entropy of the universe as a time-symmetric
first order Markov process, either a Focker-Planck or master equation process to be 
precise, depending on the state space under consideration~\citep{lawler2018introduction,serfozo2009basics}.
(The time-symmetry is necessary to reflect the time-symmetry of the microscopic laws of physics.)
Second, it stipulates that the entropy of the early universe was extraordinarily less than it is now. 
Using informal reasoning, those two assumptions 
have been taken to jointly imply that the trend of the stochastic process of the entropy of the universe
evolving into our past from our present is monotonically decreasing.

However, note that the Past Hypothesis actually involves knowing the random variable
at two times, not just one (in our case, knowing the entropy at both the present
and the distant past). It is well-known that the proper way to 
calculate the marginal distributions of a random variable evolving under a time-symmetric
Markov process given its values at two times is by using a 
``Brownian bridge''. 
In general, because the underlying stochastic process is symmetric, the Brownian bridge calculation
will lead to the conclusion that in the very recent past, just before the present, the entropy of the universe was \textit{not} 
likely to be lower than
it is today, but is actually more likely to be slightly \textit{higher}. Then as one looks further into the past from the present,
 the expected coarse-grained entropy starts
decreasing, and then falls precipitously, to reach the given, extremely low value in the distant past. 

In mesoscopic systems, with a relatively small number of degrees of freedom, the stochastic process has
enough diffusion for this ``turnover'' effect to be readily observable. The result is that the second law of thermodynamics
would be violated if one moves a very small amount into the past towards a point in time with a known, very
low value of entropy. As a result, the phenomenon that Type-3 memory systems rely on would no longer hold.

In the macroscopic system of the cosmological
universe though, one would expect the diffusion term in Markov process to be so much smaller than the drift
term, i.e., for the variance of the dynamics to be so much smaller than the underlying trend, that it would require extremely careful
and precise experiments to discern the turnover effect. Accordingly, one would suppose that Type-3 memory
systems are indeed justified in their reliance on the second law. However, it would be interesting to
calculate the precise magnitude of the turnover effect in our physical universe, to confirm this supposition.

\section*{Acknowledgement}
DHW would like to acknowledge the Santa Fe Institute for support.

\bibliographystyle{chicago}
\bibliography{bibliography}

\end{document}